\documentclass[12pt,a4paper,twoside]{article}
\usepackage{epsfig}
\usepackage{amssymb,latexsym,amsfonts}

\newcommand{\be}{\begin{equation}}
\newcommand{\ee}{\end{equation}}
\newcommand{\bd}{\begin{displaymath}}
\newcommand{\ed}{\end{displaymath}}
\newcommand{\ba}{\begin{array}}
\newcommand{\ea}{\end{array}}
\newcommand{\bt}{\begin{tabular}}
\newcommand{\et}{\end{tabular}}

\newcommand{\bea}{\begin{eqnarray}}
\newcommand{\eea}{\end{eqnarray}}
\newcommand{\bean}{\begin{eqnarray*}}
\newcommand{\eean}{\end{eqnarray*}}

\newcommand{\hlf}{\frac{1}{2}}

\newcommand{\thlf}{\frac{3}{2}}
\newcommand{\fhlf}{\frac{5}{2}}

\newcommand{\inp}[2]{\langle #1, #2 \rangle}

\newcommand{\Z}{\mathbb{Z}}

\newcommand{\R}{\mathbb{R}}
\newcommand{\C}{\mathbb{C}}

\newcommand{\mod}{\textrm{mod}}

\newcommand{\df}[1]{{} * \! #1}

\begin{document}
\begin{titlepage}

\begin{center} 
\today \hfill                  hep-th/0402090

\vskip 2 cm
{\Large \bf $E_{11}$: Sign of the times\\}
\vskip 1.25 cm
{Arjan Keurentjes\footnote{email address: Arjan@tena4.vub.ac.be}}\\
\vskip 0.5cm
{\sl Theoretische Natuurkunde, Vrije Universiteit Brussel,\\and\\
 the
  International Solvay Institutes,\\ Pleinlaan
  2, B-1050 Brussels, Belgium \\}
\end{center}
\vskip 2 cm
\begin{abstract}
\baselineskip=18pt
We discuss the signature of space-time in the context of the
$E_{11}$-conjecture. In this setting, the space-time signature depends
on the choice of basis for the ``gravitational sub-algebra'' $A_{10}$,
and  Weyl transformations connect interpretations with different signatures
of space-time. Also the sign of the 4-form gauge field term in the
Lagrangian enters as an adjustable sign in a generalized
signature. Within $E_{11}$, the combination of space-time signature 
(1,10) with conventional sign for the 4-form term, appropriate to
$M$-theory, can be transformed to the signatures (2,9) and (5,6) of
Hull's $M^*$- and $M'$- theories (as well as (6,5), (9,2) and
(10,1)). Theories with other signatures organize in orbits
disconnected from these theories. We argue that when taking $E_{11}$
seriously as a symmetry algebra, one cannot discard theories with
multiple time-directions as unphysical. We also briefly explore links
with the $SL(32,\R)$ conjecture.   
\end{abstract}

\end{titlepage}
\section{Introduction}

The discovery of duality symmetries has revolutionized the field of
high energy physics. Many characteristics of a theory turn out to
depend on ones point of view. Duality symmetries mix fields of
different types, turn local excitations of a field into solitons, and
connect theories in spaces of different topology, or even different
dimensions \cite{Hull:1994ys, Witten:1995ex}.

A fascinating, but puzzling discovery is that suitable duality
symmetries of string and M-theory can even change the signature of
space-time \cite{Hull:1998vg, Hull:1998ym, Hull:1998fh}. Although at
the classical level a theory with more than one time-direction is as
easily described as any other, it is not straightforward to extend
this to the quantum regime. Moreover, the construction of
\cite{Hull:1998vg, Hull:1998ym} involves the compactification of a
\emph{time-like} direction and therefore inevitably leads to closed
time-like curves, with all the problems associated to them. Therefore,
even though it is clear that the ideas of \cite{Hull:1998vg,
  Hull:1998ym} are correct and useful at the mathematical level, it is
not clear what physical significance should be attached to them.

It is believed that the ultimate fusion of the covariance properties
of the elf-bein of 11 dimensional supergravity \cite{Cremmer:1978km},
with the exceptional $E$-type duality symmetries of compactified
maximal supergravity \cite{Cremmer:1979up, Julia:1980gr,
  Julia:1982gx}, can be achieved within a huge Kac-Moody algebra,
which is called $E_{11}$ \cite{West:2001as}. For the $E_n$ groups with
$n<11$ that are associated to dimensional reduction to $11-n$
dimensions, we may always choose the time-direction to be transverse
to the dimensions we are reducing over (although it can be included
\cite{Hull:1998br}), but for $E_{11}$ this is no longer possible: We
have to deal with the imprint that a time direction makes on the algebra.  

The conjecture states that $E_{11}$ is realized non-linearly, as a
coset symmetry $E_{11}/H_{11}$. Technically, as we will explain in
section \ref{sig}, the signature of space-time can be incorporated by
introducing a number of signs in the denominator sub-algebra $H_{11}$
\cite{Englert:2003py, Keurentjes:2003hc} (these signs also play a
crucial role in the computations in \cite{Schnakenburg:2003qw}). The
$H_{11}$ algebra relations imply that such signs will proliferate
throughout the algebra. According to \cite{West:2002jj,
  Nicolai:2003fw, Kleinschmidt:2003mf} (see also \cite{Damour:2002cu})
the matter content of the theory can be recovered by making a level
decomposition with respect to a ``horizontal'' $A_{10}= sl(11,\R)$
sub-algebra. The signs that were introduced in the gravity sector,
imply signs in the sectors corresponding to the gauge fields, and
beyond. Dual descriptions of the theory correspond to alternative
choices for the $A_{10}$ algebra. There is no reason to expect that
all choices for $A_{10}$ will lead to the same signature of
space-time, and indeed this turns out not to be the case. The
possibility to exchange signs between the gravity and gauge sector can
be seen as the algebraic explanation for signature-changing dualities.   

This paper studies the imprint of the space-time signature (1,10) on
the $E_{11}$ algebra (throughout this paper we will denote the
space-time signature as $(t,s)$, with $t$ the number of time-like
directions, and $s$ the number of space-like directions). Our main
message will be that the version of $E_{11}/H_{11}$ 
that contains the theories with signature (1,10) also contains the
signatures (2,9), (5,6),(6,5), (9,2), and (10,1). We therefore argue
that if one takes $E_{11}$ seriously as a symmetry algebra, one cannot
discard the $M^*$- and $M'$- theories as unphysical; they come for free
with the algebra. 

In absence of certain physical requirements, such as supersymmetry,
all signatures are possible: The bosonic sector that is encoded in
$E_{11}$ does not impose any restrictions. We study how all possible
signatures connect under the Weyl group of $E_{11}$. We emphasize that
in absence of requirements as signature (1,10), or a definite sign for
the 4-form gauge-field term (requirements which we also have to
give up for M-theory as it seems), these correspond to acceptable
theories. They however cannot be supersymmetrized. 

In this paper we introduce some techniques, and work them out very
explicitly. The mathematically inclined reader will notice that these
techniques allow some level of abstraction, and are by no means
limited to $E_{11}$. In a subsequent paper \cite{Keurentjesprep} we
will apply these techniques to elucidate the full (space- and
time-like) duality web for M-theory and its cousins, to find among
other things a few duality groups that seem to have gone unnoticed in
previous works.  

\section{Definition and properties of $E_{11}$}

In this section we recall some facts about the general theory of
Kac-Moody algebra's \cite{Kac:gs}, and $E_{11}$ in particular.
We start by drawing the Dynkin diagram of $E_{11}$.

\begin{figure}[ht]
\begin{center}
\includegraphics[bb=100 650 500 720, width=10cm]{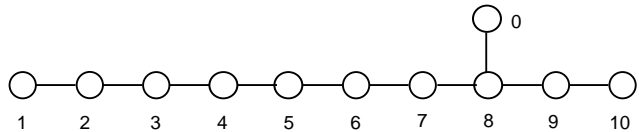}
\caption{The Dynkin diagram of $E_{11}$.}\label{e11fig}
\end{center}
\end{figure}
From this diagram the Cartan matrix $A=(a_{ij})$, with $i,j$ in the
index set $I \equiv \{0, 1, \ldots 
10 \}$, may be reconstructed by setting
\be
a_{ij} \equiv \left\{ 
\ba{rl}
2  & \textrm{if }i=j; \\
-1 & \textrm{if }i,j\textrm{ connected by a line;}\\
0  & \textrm{otherwise.}\\
\ea
\right.
\ee
The Cartan Matrix is symmetric, and $\det(A)=-2$. We choose a real
vector space ${\cal H}$ of dimension 11 and linearly independent
sets $\Pi = \{ \alpha_0, \ldots,\alpha_{10} \} \subset {\cal H}^*$
(with $\cal H^{*}$ the space dual to ${\cal H}$) and $\Pi^{\vee} = \{
\alpha_0^{\vee}, \ldots,\alpha_{10}^{\vee} \} \subset {\cal H}$,
obeying $a_{ij} = \alpha_j(\alpha_i^{\vee}) \equiv
\inp{\alpha_j}{\alpha_i^{\vee}}$. The elements of the set $\Pi$ are
called the simple roots.

From the Cartan matrix the algebra $E_{11}$ can be constructed. The
generators of the algebra consist of 11 basis elements $h_i$ for the
Cartan sub algebra ${\cal H}$ together with 22 generators
$e_{\alpha_i}$ and $e_{-\alpha_i}$ ($i \in I$), and of algebra
elements obtained by taking multiple commutators of these. These
commutators are restricted by the algebraic relations (with 
$h,h'\in{\cal H}$): 
\be
[h,h']=0 \quad [h, e_{\alpha_j}] = \inp{\alpha_{j}}{h} e_{\alpha_j}
\quad [h,e_{-\alpha_j}] = -\inp{\alpha_j}{h} e_{-\alpha_j} \quad
       [e_{\alpha_i}, e_{-\alpha_j}] =\delta_{ij} \alpha_i^{\vee}; 
\ee
and the Serre relations
\be
\textrm{ad}(e_{\alpha_i})^{1-a_{ij}} e_{\alpha_j} =0, \qquad
\textrm{ad}(e_{-\alpha_i})^{1-a_{ij}}e_{-\alpha_j} =0.  
\ee
Implicit in the conjecture is that one is working with a real form of
$E_{11}$, to be precise the split real form, which is generated by
linear combinations of the above generators with \emph{real} coefficients.

As is well known, the Cartan matrix of $E_{11}$, implies that the root
space is of signature $(10,1)$. Consider the space defined by
11-tuples forming vectors $p = (p_1, \ldots p_{11}) \in \R^{11}$, with norm
\be
 ||p||^2 = \sum_{i=1}^{11} p_i^2 - \frac{1}{9} \left(\sum_{i=1}^{11}
 p_i \right)^2,
\ee
and inner product $\inp{ \, }{}$ defined by the norm via
\be \label{inp}
\inp{a}{b} = \frac{1}{2} \left( ||a +b||^2 -||a||^2 - ||b||^2
\right)=\sum_{i=1}^{11} a_i b_i - \frac{1}{9} \left(\sum_{i=1}^{11}
 a_i \right)\left(\sum_{i=1}^{11}
 b_i \right).
\ee
With this inner product, the signature of the root space is (10,1):
the 10 dimensional subspace of vectors $x$ defined by $\sum_{i=1}^{11}
 x_i =0$ contains only vectors of positive norm; the 1 dimensional orthogonal
 complement consisting of vectors $x$ of the form $x_i = \lambda,
 \lambda \in \R, \forall i$ has vectors of negative norm (these
 choices were inspired by the choice of metric on the $E_{10}$ root
 space in \cite{Damour:2002et, Brown:2004jb}). 
 
We realize the simple roots of $E_{11}$ in the above space as
\bea
\alpha_i & = & e_i - e_{i+1}, \qquad i = 1, \ldots, 10;\\
\alpha_0 & = & e_9 + e_{10} + e_{11}.
\eea
Here $(e_i)_j = \delta_{ij}$, and the reader should notice that with
the inner product (\ref{inp}) these are \emph{not} unit vectors. The
root lattice of $E_{11}$, which we will call $P_{11}$, consists of
linear combinations of the simple roots, with coefficients in
$\Z$. This lattice can be characterized as 
consisting of those vectors $a$ whose components $a_i$ are integers,
and sum up to three-folds $3k$. The integer $k$ counts the occurrences
of the ``exceptional'' root $\alpha_0$, and for roots equals the level
as defined in \cite{West:2002jj, Nicolai:2003fw} (see also
\cite{Damour:2002cu}). 

Via the innerproduct defined as above, and the relations $a_{ij} =
\inp{\alpha_i}{\alpha_j^{\vee}}$ it follows that there is a natural
bijection $\alpha_i \rightarrow \alpha_i^{\vee}$, in which the
components of each $\alpha_i^{\vee}$ turn out to be identical to those of
$\alpha_i$ (This is because $E_{11}$ is a simply-laced algebra). The
coroot lattice, consisting of linear combinations with integer
coefficients of $\alpha_i^{\vee}$ may therefore be identified with the
root lattice. In view of this, our two different definitions of $\inp{
  \, }{}$ are numerically equivalent.
 
There is the root space decomposition with respect to the Cartan
subalgebra, $E_{11} = \oplus_{\alpha \in {\cal H}^*} g_{\alpha}$, with 
\be
g_{\alpha} = \{ x \in E_{11}: [h,x] = \inp{\alpha}{h} x \ \forall h
\in {\cal H} \} 
\ee
The set of roots of the algebra, $\Delta$, are defined by
\be
\Delta = \{ \alpha \in {\cal H}: g_{\alpha} \neq 0, \alpha \neq 0  \}
\ee
The roots of the algebra form a subset of the root lattice $\Delta
\subset P_{11}$. The set of positive roots $\Delta^+ \subset \Delta$
is the subset of roots whose expansion in the simple roots
involves non-negative integer coefficients only. We denote the
basis elements of $g_{\alpha}$ by $e^k_{\alpha}$, where $k$ is a
degeneracy index, taking values in $\{1, \ldots, \dim(g_{\alpha})
\}$. If $\dim(g_{\alpha}) =1$, we will drop the degeneracy index, and
write $e_{\alpha}$ for the generator. This is in accordance with previous
notation, as $\dim(g_{\alpha_i}) =1$ when $\alpha_i$ is a simple
root. By using the Jacobi identity, one can easily prove that
$[e^i_{\alpha}, e^j_{\beta}] \in g_{\alpha+\beta}$, if this commutator
is different from zero.  

The inverse Cartan matrix is given by (notice the sign in front):
\be
-A^{-1}= \frac{1}{2}\left( \ba{rrrrrrrrrrr}
-1 & 0 & 1 & 2 & 3 & 4 & 5 & 6 & 4 & 2 & 3 \\
 0 & 0 & 2 & 4 & 6 & 8 & 10& 12& 8 & 4 & 6 \\
 1 & 2 & 3 & 6 & 9 & 12& 15& 18& 12& 6 & 9 \\
 2 & 4 & 6 & 8 & 12& 16& 20& 24& 16& 8 & 12\\
 3 & 6 & 9 & 12& 15& 20& 25& 30& 20& 10& 15\\
 4 & 8 & 12& 16& 20& 24& 30& 36& 24& 12& 18\\
 5 & 10& 15& 20& 25& 30& 35& 42& 28& 14& 21\\
 6 & 12& 18& 24& 30& 36& 42& 48& 32& 16& 24\\
 4 & 8 & 12& 16& 20& 24& 28& 32& 20& 10& 16\\
 2 & 4 & 6 & 8 & 10& 12& 14& 16& 10& 4 & 8 \\
 3 & 6 & 9 & 12& 15& 18& 21& 24& 16& 8 & 11\\
\ea \right)
\ee

The fundamental coweights $\omega_i$ are defined by 
\be
\inp{\alpha_i}{\omega_j} = \delta_{ij}.
\ee
They can be expressed in the coroots as $\omega_i =
(A^{-1})_{ij} (\alpha_j)^{\vee}$. In the basis we have chosen, the
explicit form of the fundamental coweights is: 
\be
\ba{r@{=(}r@{,\ }r@{,\ }r@{,\ }r@{, \ }r@{, \ }r@{, \ }r@{, \ }r@{, \
  }r@{, \ }r@{, \ }r@{ \ )}}
-\omega_1 & -\hlf &  \hlf&  \hlf &  \hlf & \hlf & \hlf&
 \hlf& \hlf& \hlf& \hlf& \hlf \\ 
-\omega_2 & 0 & 0 &  1& 1& 1& 1& 1& 1& 1& 1& 1\\
-\omega_3 &  \hlf &  \hlf &  \hlf &  \frac{3}{2} &  \frac{3}{2} &
 \frac{3}{2} &  \frac{3}{2} &  \frac{3}{2} &  \frac{3}{2} &
 \frac{3}{2} &  \frac{3}{2} \\
-\omega_4 &  1&  1&  1&  1&  2&  2&  2&  2&  2&  2&  2\\
-\omega_5 &  \frac{3}{2}&  \frac{3}{2}&  \frac{3}{2}&  \frac{3}{2}&
 \frac{3}{2}&  \frac{5}{2}&  \frac{5}{2}&  \frac{5}{2}&  \frac{5}{2}&
 \frac{5}{2}&  \frac{5}{2} \\
-\omega_6 &  2&  2&  2&  2&  2&  2&  3& 3& 3& 3& 3 \\
-\omega_7 &  \frac{5}{2} &  \frac{5}{2} &  \frac{5}{2} &  \frac{5}{2}
&  \frac{5}{2} &  \frac{5}{2} &  \frac{5}{2} &  \frac{7}{2}&
 \frac{7}{2}&  \frac{7}{2}&  \frac{7}{2}\\
-\omega_8 &  3 &  3 &  3 &  3 &  3 &  3 &  3 &  3 &  4 &  4 &  4\\
-\omega_9 &  2 &  2 &  2 &  2 &  2 &  2 &  2 &  2 &  2 &  3 &  3 \\
-\omega_{10} &  1 &  1 &  1 &  1 &  1 &  1 &  1 &  1 &  1 &  1 & 2\\
-\omega_0 & \frac{3}{2} &  \frac{3}{2} &  \frac{3}{2} &  \frac{3}{2}
&  \frac{3}{2} &  \frac{3}{2} &  \frac{3}{2} &  \frac{3}{2} &
\frac{3}{2} &  \frac{3}{2} &  \frac{3}{2} \\ 
\ea 
\ee

The coweight lattice, which we call $Q_{11}$ consist of linear
combinations of the fundamental coweights with coefficients in
$\Z$. It can be characterized as those vectors that have components
that are either all integers, or all odd integers 
divided by 2, and whose components sum up to multiples of
$\frac{3}{2}$. It is clear that the coweight lattice contains the coroot
lattice, which is an index 2 sublattice. 

We will make much use of the Weyl group $W_{11}$ of $E_{11}$. This is
the group generated by the Weyl reflections $w_i$ in the simple roots, 
\be
w_i(\beta) = \beta - 2
\frac{\inp{\alpha_i}{\beta}}{\inp{\alpha_i}{\alpha_i}} \alpha_i 
\ee
The Weyl group leaves the inner product invariant
\be \label{winp}
\inp{w(\alpha)}{w(\beta)}=\inp{\alpha}{\beta} \qquad w \in W_{11}
\ee

The Weyl group includes reflections in the non-simple roots. In the
basis we have chosen the Weyl reflections $w_i$ in $\alpha_i$,
$i=1,\ldots,10$ permute entries of the row of 
11 numbers. Crucial to our story are the Weyl reflections in
$\alpha_{0}$, and other roots that have $\alpha_{0}$ exactly once in
their expansion. We will need these often, and find it convenient
to introduce specific notation to represent these roots. They are all
of the form 
\bd
\beta_{ijk} = e_i + e_j + e_k, \qquad i<j<k
\ed
For example $\alpha_{0} = \beta_{9 \, 10 \, 11}$

\section{Space-time signature and U-duality algebra} \label{sig}

After the many technicalities introduced in the previous section, we
proceed with explaining in what way they enter the discussion on
physics. 

\subsection{Compact and non-compact generators of $H_{11}$} \label{h11}

The conjectured $E_{11}$-symmetry of $M$-theory is non-linearly
realized \cite{West:2001as}. The relevant variables are not described
by $E_{11}$, but by the coset $E_{11}/H_{11}$. By analogy to the
cosets appearing in dimensional reduction (see
e.g.\cite{Cremmer:1999du, Keurentjes:2002xc} and also
\cite{Damour:2002cu}), it is proposed that the degrees of freedom  can
be retrieved from $E_{11}$ by decomposing the algebra with respect to
a horizontal $A_{10}$ algebra \cite{West:2002jj,
  Nicolai:2003fw, Kleinschmidt:2003mf}. The real form corresponding to
$A_{10} = SL(11,\R)$; the $11^{th}$ generator of the Cartan subalgebra 
(generating a non-compact scaling symmetry isomorphic to $\R$) can be
added to this to give the algebra of $GL(11,\R)$. This is argued to be
the $GL(11,\R)$ relevant to the local vielbein, which can be regarded
as an element of $GL(11,\R)/SO(1,10)$. Hence, when specifying the real form
of $H_{11}$, we will start with the horizontal subalgebra $SO(1,10)$. 

The maximal compact subgroup inside $E_{11}$ can be defined as the
group invariant under the Cartan involution; it consists of generators
of the form $e^k_{\alpha} -e^k_{-\alpha}$. In our conventions
$(e^k_{\alpha})^{\dag} = e^{k}_{-\alpha}$, and therefore $e^k_{\alpha}
-e^k_{-\alpha}$ is anti-hermitian, implying that it generates a compact
symmetry. It was observed in \cite{Englert:2003py, Keurentjes:2003hc}
(see also \cite{Schnakenburg:2003qw} for an application)
that the possibility of non-compact denominator groups can be taken
into account by including a sign $\epsilon_{\alpha}= \pm 1$ in the
generator, such that the real form of $H_{11}$ will be generated by
generators of the form  
\be
T^k_{\alpha} = e^k_{\alpha} - \epsilon_{\alpha} e^k_{-\alpha}, \qquad
\alpha \in \Delta^+
\ee
Because both $\alpha$ and $-\alpha$ enter in the definition, we can
restrict to $\alpha \in \Delta^+$. We will explain in a
moment that $\epsilon_{\alpha}$ cannot depend on the degeneracy index
$k$, but we have already omitted it from our notation. 

For $\epsilon_{\alpha} =1$ $T^k_{\alpha}$ is anti-hermitian, whereas
for $\epsilon_{\alpha} = -1$ it is hermitian. Now it is a simple fact
that for $H,H'$ hermitian, and $A,A'$ anti-hermitian, 
\bd
[H,H'], \textrm{ and } [A,A']
\ed
are anti-hermitian, whereas
\bd
[H,A]
\ed
is hermitian. Using that the structure constants of $E_{11}$, and
hence of $H_{11}$ are real, we deduce that the hermiticity
properties of the generators $T^k_{\alpha \pm \beta}$ follow directly
from those of $T^k_{\alpha}$ and $T^k_{\beta}$. Upon using that all
elements $T^k_{\alpha}$ can be formed from taking multiple commutators
of the $T_{\alpha_i}$, with $\alpha_i$ the simple roots (where we can
again drop the degeneracy index $k$), we see that $\epsilon_{\alpha}$
does depend on $\alpha$, but cannot depend on the degeneracy index.  

We now form a $\Z_2$-valued linear function $f(\alpha)$ on the root
lattice, by defining for the simple roots
\bea
f(\alpha_i) = 0 & & \textrm{if }(T_{\alpha_i})^{\dag} = -T_{\alpha_i} \\
f(\alpha_i) = 1 & & \textrm{if }(T_{\alpha_i})^{\dag} = T_{\alpha_i}
\eea
and continue on the root lattice by linearity. If $\alpha$ belongs to
the root lattice of $E_{11}$, then $f(\alpha)$ specifies whether the
corresponding generator(s) $T^i_{\alpha}$ of $H_{11}$ are compact or
non-compact generators: $f(\alpha) = 0$ implies anti-hermiticity, and therefore
compactness of the symmetry generated by this generator, while
$f(\alpha)=1$ implies hermiticity, and non-compactness of the
generator. Consequently, the function $f$ specifies the real form of
the denominator subgroup $H_{11}$. 

It is important to realize that, even though $H_{11}$ consists of
infinitely many elements, that there are only finitely many
$\Z_2$-valued linear functions on the root lattice. These are completely
specified by their values on the simple roots, so there are only
$2^{11}=2048$ such functions. It is actually easy to describe them, they are
all of the form
\be
f(\alpha) = \sum_i p_i \inp{\alpha}{\omega_i} \ \mod \ 2, \qquad \alpha
\in P_{11},
\ee
where, as $\inp{\alpha}{\omega_i} \in \Z$, the coefficients $p_i$ can
be chosen to be in $\Z_2$. Below we will specify the explicit entries
of $f$ as a row of 11 numbers. The $\mod \ 2$ property of the function
$f$ then translates to equivalence under shifts by elements of
$2Q_{11}$ (which is a sublattice of the coroot lattice, which we have
identified with the root lattice $P_{11}$).   

The function $f$ is completely specified by its values on a basis of
simple roots. There are however many such bases. We regard two
choices of basis for the root lattice as equivalent if they are
related by a sequence of Weyl reflections, that is by an element of
the Weyl group. If the original basis is given by $\{ \alpha_i \}_{i
  \in I}$, then the Weyl transformed basis is given by $\{ w(\alpha_i)
\}_{i \in I}, \ w \in W_{11}$. The transformation $w$ does not alter
$H_{11}$, and hence at the mathematical level describes the same
theory. It may however rotate signs from the gravity into the gauge
sector, and vice versa, and hence the interpretation of the
mathematical theory may result in a different space-time signature
and/or a different sign in front of the 4-form term. 

As, due to the property (\ref{winp}) of the Weyl group
\be
\inp{w(\alpha_i)}{f} = \inp{\alpha_i}{w^{-1}(f)}, \qquad w, w^{-1} \in W_{11},
\ee 
the theory with the Weyl rotated basis and function $f$, gives the
same theory as the original basis with a Weyl rotated $f$. It is
however easier to study the action on one linear function $f$ than on
a basis of 11 roots $\alpha_i$. Finding all possible non-compact forms
that can appear in the denominator group therefore amounts to
classifying all functions $f$ up to Weyl reflections. 

We have transcribed our problem to a mathematical setting. Before
turning to $E_{11}$ we need to sort one more thing out: The relation
between the function $f$ and the signature of space-time.   

\subsection{Space-time signature from $H_{11}$}

To figure out what the space-time signature corresponding to a certain
function $f$ is, we have to study its values on the $A_{10}$
sub-algebra that defines the gravitational sector of the theory. Again
it will suffice to study its value on the simple roots, but it is
instructive to check the action on the full set of roots (which for
$A_{10}$, in contrast to $E_{11}$, is a finite set, and therefore
manageable). 

The 55 positive roots corresponding to the $A_{10}$ subalgebra are given by
\be
 \sum_{i=k}^l \alpha_i, \qquad 1 \leq k \leq l\leq 10.
\ee
It will be clear that $f$ and $f+\omega_0$ correspond to the same
signature of the $A_{10}$ algebra, so for the time being we will
ignore $\omega_0$. We will return to its significance later.

Consider $f=\omega_1$. This gives $1$ on all roots having $\alpha_1$
in its expansion. More precisely, it gives $1$ on all roots in which
the root $\alpha_1$ appears an odd number of times, but in all roots
of $A_{10}$ it appears at most once. There are 10 positive roots of
$A_{10}$ containing $\alpha_1$, and 45 which do not contain it. This
means that 10 out of the $T_{\alpha}$ generators, will be non-compact
generators, while the other 45 are compact ones. Knowing that the
algebra generated by the $T_{\alpha}$ is a real form of $so(11,\C)$,
we immediately identify the algebra as the one of $so(10,1)$.

As another example, set $f = \omega_2$. The same procedure as above
reveals that there are 18 non-compact generators, and  37 compact
ones. Hence the algebra must be $so(2,9)$. Similarly, it is easy to verify
that $f=\omega_p$ corresponds to $so(p,11-p)$.

To consider a more complicated example, take $f = \omega_1 +
\omega_3$. There are 10 roots containing $\alpha_1$, and $24$
containing $\alpha_3$. There are however 8 roots that contain
$\alpha_1$ as well as $\alpha_3$, and that are therefore mapped to $0 \
\mod  \ 2$ by $f$. There are therefore $18$ non-compact generators, and
again the denominator sub-algebra is $so(2,9)$. As a matter of fact,
the coweight $\omega_1 + \omega_3$ is equivalent to
$\omega_2$, under a suitable combination of Weyl reflections and
translations over elements from $2Q_{11}$.

A less abstract and more direct way to see the group is the
following. Let $(x_1, x_2, \ldots, x_{11})$ be the coordinates on the
eleven dimensional tangent space (on which the denominator subgroup of
$A_{10}$ acts). Take $T_{\alpha_i}$ ($1 \leq i \leq 10$) to be the
generator that mixes the $x_i$ and $x_{i+1}$. Now $T_{\alpha_i}$
generates $SO(2)$ if it is anti-hermitian, and $SO(1,1)$ if it is
hermitian. Therefore if $f(\alpha_i)=0$, then $x_i$ and $x_{i+1}$
correspond both to space-, or both to time-like directions, whereas
when $f(\alpha_i)=1$, one of them is a space-like and the other one a
time-like direction. Evaluating $f$ successively on $\alpha_{10}$,
$\alpha_9$, etc. immediately gives the signature.
 
This can be summarized in the following simple procedure: To compute
the denominator group in $A_{10}$ corresponding to $f$, write $f$ as
the sum of $n$ weights, and order these as
\bd
\omega_{k_1}+ \omega_{k_2} + \ldots + \omega_{k_n}, \qquad 10 \geq k_1 > k_2 >
\ldots > k_n \geq 0.
\ed
Then the algebra is given by 
\be
so(k,11-k); \qquad k = \sum_{i=1}^{n} (-1)^{i+1} k_{i}. 
\ee
Note that $k \geq 0$, and $\omega_0$ does not contribute.

Having found that the signature of space-time is encoded in the
weights $\omega_1,\ldots \omega_{10}$, what is the significance of
$\omega_0$ ? There is yet another source of minus signs in the
algebra, coming from the sign in front of the $\df{G} \wedge G$ term
in the 11-dimensional Lagrangian, corresponding to the $3$-form gauge
field. A certain choice of space-time signature implies that there are
all kinds of signs implicit in the contraction of this term. We can
however also adjust the overall sign in front of this term. It is
precisely this minus-sign that is encoded in $\omega_0$. That it works
this way can actually be easily seen by comparing to the sigma-model
action for the effective theories representing toroidal
compactification of less than 11 dimensions \cite{Cremmer:1997ct,
  Keurentjes:2002xc}, but we will discuss this more completely
elsewhere \cite{Keurentjesprep}. 

We define a ``generalized signature'' $(t,s,\pm)$, denoting by $t$ the
number of time directions, by $s$ the number of space-like direction,
and $\pm$ giving the sign in front of the 4-form gauge-field term,
relative to the conventional one. In terms of this sign, the bosonic
part of the 11 dimensional Lagrangian equals
\be
R - (\pm \df{G} \wedge G),
\ee
plus the Chern-Simons term, that is of no relevance in this
discussion, as its sign can be changed by redefining the 3-form
potential $C_{(3)} \rightarrow -C_{(3)}$, which is a transformation without any
relation to space-time.

The discussion in the above gives the relative signature $|t-s|$, we
cannot distinguish between signature $(t,s)$ and $(s,t)$. The ${11
  \choose 3} =165$ combinations $T_{\alpha}= e_{\alpha}
-\epsilon_{\alpha} e_{-\alpha}$, for which $\alpha_0$ enters with
multiplicity 1 in the expansion of $\alpha$, are loosely interpreted
as associated to the components of the 3-form potential giving rise to
the 4-form field strength \cite{West:2002jj, Nicolai:2003fw}. In a
space-time with signature $(t,s)$, the number of components of the 3
form with an odd number of space-like directions is   
\be \label{odd}
{t \choose 0}{s \choose 3} + {t \choose 2}{s \choose 1}
\ee
(where we put binomials ${p \choose q}$ that would be ill-defined,
that is when $q > p$, to zero). These correspond to compact generators. On
the other hand, the number of components of the 3-form with an even
number of space-like dimensions (giving non-compact generators), is 
\be \label{even}
{t \choose 1}{s \choose 2} + {t \choose 3}{s \choose 0}.
\ee
Interchanging $t \leftrightarrow s$ interchanges (\ref{odd}) with
(\ref{even}). Hence, if we denote a function $f$, representing $t$
time directions, $s$ space-directions, and a conventional sign in
front of the 4-form term by $f_{(t,s,+)}$, then there is an equality  
\be
f_{(t,s,+)} = f_{(s,t,-)}
\ee
where $f_{(s,t,-)}$ is a function representing a theory with $s$ time
directions, $t$ space-directions, and an unconventional sign in front
of the 4-form term. 

Theories with generalized signature $(t,s,+)$ and $(s,t,-)$ exist, and
are described by the same function. To argue that they can be
connected by Weyl transformations, we must look at the
tangent space with coordinates $(x_1, \ldots, x_{11})$ and work out the
corresponding relative signature. Then we \emph{choose} the absolute
signature, by assigning that one group of coordinates describes
space-like, and the other group time-like directions. If we now fix a
group of space-like (time-like) coordinates, and act with Weyl
transformations on generators that preserve these coordinates, we can
reach other theories, where the group of fixed coordinates still
describes space-like (time-like) directions. Doing this in successive
steps often (but not always) the theory with space-time signature
$(s,t)$ can be reached from the theory with signature $(t,s)$. This is
essentially the algebraic transcription of the procedure followed in
\cite{Hull:1998ym} to deduce the existence of the theory with
space-time signature $(s,t)$ from the one of signature $(t,s)$, as in
the toroidal context, the Weyl reflection in $\beta_{ijk}$ corresponds
to choosing a 3-torus, and rewriting to the theory for the dual 3-torus. 

\section{$E_{11}$ and the signature of space-time}

By itself, $E_{11}$ does not prefer any signature of space-time. As
should have become clear in the previous, we are free to choose any
space-time signature, as well as the overall sign for the 4-form
term in the Lagrangian. We can, and will work out all possible
signatures and theories, but emphasize that only a subset of them
meets certain physical requirements, such as compatibility with supersymmetry.

\subsection{Space-time signatures compatible with $M$-theory}

We will start with the algebra that represents the bosonic sector of
the conventional 11 dimensional supergravity \cite{Cremmer:1978km},
having signature $(1,10)$ and a conventional sign in front of the 4-form term.

Such a theory can be represented by the function $f_{(1,10,+)}$, given by 
\be
f_{(1,10,+)} = f_{(10,1,-)}= \left(-\hlf , \hlf , \hlf, \hlf, \hlf,
\hlf, \hlf, \hlf, \hlf, \hlf, \hlf \right)  
\ee
which is actually $-\omega_1$, but under shifts by $2Q_{11}$ we do not need
to worry about the overall sign. There are ${11 \choose 1} =11$
permutations of the entries, which give different functions, but do of
course correspond to the same space-time signature (and 4-form
term).

To find another space-time signature, described by the same
non-compact form of $H_{11}$, we apply a Weyl reflection in
$\beta_{123}$. After permuting the entries we arrive at
\be
f_{(2,9,-)}= f_{(9,2,+)} = \left( \thlf , \thlf , \hlf, \hlf, \hlf,
\hlf, \hlf, \hlf, \hlf, \hlf, \hlf \right)
\ee
By permutations there are ${11 \choose 2} = 55$ choices that give the
same space-time signature.

We have set the third index on $f_{(2,9,-)}$ to $(-)$. If we assume a
space-time signature $(2,9)$ the $165$ components of the 3-form
gauge-field divide into 93 terms with positive sign, giving compact
generators, and 72 of the opposite sign, giving non-compact
generators. It is however easily verified that $f_{(2,9,-)}$ results
in 72 compact generators in the 3-form sector, and 93 non-compact
ones. Therefore we have to invoke the extra minus sign in front of the
4-form field strength to make the correspondence work. Note that
this extra minus sign is also crucially present in $M^*$-theory
\cite{Hull:1998ym}. 

To find yet another space-time signature, we apply a Weyl reflection
to $f_{(2,9,-)}$ in $\beta_{345}$, to arrive at
\be \label{f56}
f_{(5,6,+)}= f_{(6,5,-)} = \left(\thlf , \thlf , \thlf, \thlf, \thlf,
\hlf, \hlf, \hlf, \hlf, \hlf, \hlf\right)
\ee
By permuting the entries there are ${ 11 \choose 5} = 462$ choices
that give the same space-time signature. Again the reader should take
notice of the $+$ appearing on $f_{(5,6,+)}$, this time the signs in
the 4-form field terms are in accordance with the space-time signature
(which again agrees with the findings of \cite{Hull:1998ym}).

Now let (\ref{f56}) correspond to a theory for which the coordinates
$x_1,\ldots,x_4$ correspond to time-like directions, whereas the
coordinates $x_8, \ldots, x_{11}$ correspond to space-like
directions. Then the space-time signature is $(5,6)$. Weyl reflecting
in $\beta_{567}$ and making suitable shifts over $2Q_{11}$, one
arrives at
\be
\left(\hlf , \hlf , \hlf, \hlf, \thlf,
\hlf, \hlf, \thlf, \thlf, \thlf, \thlf\right),
\ee
which is, up to permutations of its entries equivalent to
$f_{(5,6,+)}$, but under the identifications we have made for $x_1,
\ldots, x_4, x_8, \ldots, x_{11}$ it must represent a theory with
space-time signature $(6,5)$. 

Continuing from here we can reach all of the signatures $(1,10)$,
$(2,9)$, $(5,6)$, $(6,5)$, $(9,2)$, $(10,1)$. Further Weyl reflections
to $f_{(1,10,+)}$, $f_{(2,9,-)}$ and $f_{(5,6,+)}$ do not generate new
functions; all are, up to shifts by $2Q_{11}$ and permutation of the
entries equivalent to $f_{(1,10,+)}$, $f_{(2,9,-)}$ and
$f_{(5,6,+)}$. We conclude that these $11+55+462 = 528$ $\Z_2$-valued
functions on the $E_{11}$ root lattice form a closed orbit under Weyl
reflections. They correspond to choices of signs relevant to $M$-,
$M^*$- and $M'$-theory \cite{Hull:1998ym}. We note that it is
essentially the Weyl reflection in the exceptional root $\alpha_0$
that allows for space-time signature changing transformations.   

\subsection{Other space-time signatures}

There are three more, inequivalent orbits of the Weyl group, acting on
the $\Z_2$-valued functions defining various space-time signatures. If
we accept the existence of theories with exotic space-time signatures,
and wrong signed 4-form terms (as the $E_{11}$ conjecture, and
time-like T-duality seem to impose on us), then these other theories
are acceptable too. They however cannot be supersymmetrized, as
spinors in these signatures do not have the right number of components
to produce a supersymmetry algebra (see \cite{Hull:1998ym}).

A theory that is straightforwardly interpreted is the one
corresponding to
\be
f_{(0,11,+)} = f_{(11,0,-)} = 0,
\ee
corresponding to the compact form of the denominator algebra
$H_{11}$. It is clear that this single entry fills out a whole orbit
by itself. The ``physical'' interpretation gives a theory on a
Euclidean space-time, with the conventional sign for the 4-form
gauge field. This version may have its usual importance as
representing the symmetries of the Wick-rotated $(1,10)$ theory. There
is also a version with time-like directions only, and an unconventional
sign of the 4-form gauge-field term. This theory can\emph{not} be
reached from the Euclidean theory. 
 
We move on to the next class of theories. A convenient starting point
is the algebraic realization of the bosonic sector of 11-d
supergravity, on a Euclidean space-time, but now with the wrong sign
on the 4-form term in the action. From our previous discussion it
should be clear that this theory can represented by
\be
f_{(0,11,-)} = f_{(11,0,+)}= -\omega_0 = 
\left(\thlf,\thlf,\thlf,\thlf,\thlf,\thlf,\thlf,\thlf,\thlf,\thlf,\thlf\right).
\ee

After a Weyl reflection in $\beta_{123}$ this turns into
\be
f_{(3,8,+)} = f_{(8,3,-)} =
\left(\fhlf,\fhlf,\fhlf,\thlf,\thlf,\thlf,\thlf,\thlf,\thlf,\thlf,\thlf
\right).     
\ee
There are ${11 \choose 3}=165$ permutations that give the same
space-time signature.

A Weyl reflection in $\beta_{145}$, permutations of the entries, and
subsequent shifts over $2Q_{11}$ take us to
\be
f_{(4,7,-)}= f_{(7,4,+)}= \left( \hlf, \hlf, \hlf, \hlf,
-\hlf,-\hlf,-\hlf,-\hlf,-\hlf,-\hlf,-\hlf \right),
\ee
which gives, by permutation of the entries, ${11 \choose 4} = 330$
choices with the same space-time signature.

It is a simple exercise to show that theories of signature $(4,7)$ and
$(7,4)$ are connected by Weyl reflections. Further Weyl reflections
always take us back to $f_{(11,0,-)}, f_{(3,8,+)}$ and $f_{(4,7,-)}$,
and functions equivalent to these. Hence this orbit is closed, and
contains $1 + 165 + 330=496$ $\Z_2$-valued functions. It is amusing to
see that it precisely contains those space-time signatures not allowed
by 11-dimensional supersymmetry.  

It turns out there is one remaining orbit. A convenient starting point
is the theory with signature $(1,10)$, but the wrong sign for the
4-form term. This is described by
\be
f_{(1,10,-)} = f_{(10,1,+)} = -\omega_{10} =\left(1,1,1,1,1,1,1,1,1,1,2\right).
\ee

A Weyl reflection in $\beta_{123}$ takes us to
\be 
f_{(4,7,+)} = f_{(7,4,-)} = \left(2,2,2,2,1,1,1,1,1,1,1\right).
\ee

A Weyl reflection in $\beta_{156}$, followed by permutation of the
entries and a suitable shift over $2Q_{11}$ takes us to
\be
f_{(5,6,-)} = f_{(6,5,+)} = \left(1,1,1,1,1,1,0,0,0,0,0 \right).
\ee

A Weyl reflection over $\beta_{456}$ takes us to
\be
f_{(8,3,+)} = f_{(3,8,-)}= \left( 1,1,1,0,0,0,0,0,0,0,0 \right).
\ee

And a final Weyl reflection in $\beta_{234}$ followed by suitable
permutations takes us to
\be
f_{(9,2,-)} = f_{(2,9,+)} =\left( 1,-1,0,0,0,0,0,0,0,0,0 \right).
\ee
Notice that this function allows $110$ permutations of the entries,
but that as $f$ and $-f$ represent the same signs, these only
correspond to ${11 \choose 2} =55$ $\Z_2$-valued functions. This last
orbit contains ${11 \choose 1} + {11 \choose 2} + {11 \choose 3} + {11
\choose 4} + {11 \choose 5} =1023$ $\Z_2$-valued functions. 

Detailed examination of specific signatures reveals that the theories with
space-time signatures $(1,10)$, $(4,7)$, $(5,6)$, $(8,3)$ and $(9,2)$
are connected by Weyl reflections. Also the theories with signatures
$(10,1)$, $(7,4)$, $(6,5)$, $(3,8)$ and $(2,9)$ are connected, but it
is impossible to reach say $(1,10,-)$ from $(10,1,+)$. Hence,
like the function specifying the Euclidean theory, this single orbit
of functions actually specifies 2 orbits of theories.

We have now listed one representative for each possibility for the
generalized space-time signature. Moreover, as 
\bd
528 + 1 + 496 + 1023 = 2048,
\ed
we have succeeded in dividing all $\Z_2$-valued functions on the root
lattice of $E_{11}$ into 4 orbits of its Weyl group. As we have
explained these correspond to 6 groups of theories. These results imply
that there are 4 possible real forms of $H_{11}$ that can appear in
$E_{11}/H_{11}$, of which one corresponds to choices of signs
appropriate for $M$-, $M^*$- and $M'$-theories.

\section{Relation to the $SL(32)$-conjecture}

In \cite{Hull:2003mf} (see also \cite{Barwald:1999is}) it was observed
that the $sl(32,\R)$ algebra gives a formal symmetry of the 11-dimensional
supercovariant derivative, and it was argued that this algebra should
provide a useful tool in the classification of solutions. Proposals to
elevate $SL(32,\R)$ to a local symmetry of $M$-theory however turn out
to be problematic with the representations present in $3$-dimensional
supergravity \cite{Keurentjes:2003yu}, and appears to be incompatible
with the most general form of the $E_{11}$ conjecture, as the local
subgroup in $E_{11}$ does not have an $SL(32,\R)$ subgroup
\cite{West:2003fc, Keurentjes:2003hc}.

There is however a remarkably simple relation  between $E_{11}$ and
$SL(32,\R)$, which was outlined in \cite{West:2003fc} and completed and
generalized in \cite{Keurentjes:2003hc}. The algebra
$sl(32,\R)$ can be obtained by truncating the local subalgebra
$H_{11}$ in a specific way. The procedure essentially amounts to
truncating the $E_{11}$ algebra to exclude the generators for
which the absolute value of the level $k$ exceeds 4, and then removing
all generators from the remaining ones in $H_{11} \subset E_{11}$ that
do not fall into antisymmetric tensor representations of the
horizontal subalgebra $so(1,10)$, which is a real form of $B_5$. 

Without further discussion on the validity and merit of the
$SL(32,\R)$ conjectures, we observe that our arguments as developed
for $E_{11}$ immediately allow an easy determination of the subgroup
surviving the above truncation procedure. The generators at level 0
form the adjoint $\mathbf{55}$ of $B_5$, the generators at level $\pm
1$ combine into the 3-form $\mathbf{165}$, the generators at level
$\pm 2$ organize in a 6-form $\mathbf{462}$, of the generators at
level $\pm 3$ a 7-form $\mathbf{330}$ survives, and of the generators at level
$\pm 4$ a 10-form $\mathbf{11}$ remains, for a total of 1023
generators, which make up the Lie algebra of $A_{31}$ \cite{Hull:2003mf}. 

We have to divide these into groups of hermitian and anti-hermitian
generators. As these correspond to non-compact and compact generators,
subtracting the number of the anti-hermitian generators from the number
of the hermitian generators immediately gives the
signature of the real form of $A_{31}$ (the reader should not confuse
the signature of the algebra with the space-time signature of the
corresponding theory). 

It is perhaps most convenient to start with the Euclidean
theory, with conventional sign for the 4-form. The truncation as
described above leads to 1023 compact generators, hence a real form of
$A_{31}$ with signature $-1023$. The corresponding algebra must be $su(32)$. 

The theory of space-time signature $(1,10)$, with conventional sign
for the 4-form gauge field gives 527 hermitian generators, against 496
compact ones. Therefore the signature of the algebra is 31,
reproducing the well-known $sl(32,\R)$. 

The theory in Euclidean signature but with wrong signed 4-form term
gives anti-hermitian generators for the generators from level 0, 2 and
4, and hermitian for the ones from level 1 and 3. This gives 528
compact generators, against 495 non-compact ones, and fixes the
signature of the algebra to $-33$. We therefore identify the algebra
as the one of $su^*(32)$. 

The theory of space-time signature $(1,10)$ with unconventional sign
for the 4-form term results in 511 compact generators, versus 512
non-compact ones, giving the value 1 for the signature of the algebra, which
identifies the algebra as $su(16,16)$. 

We note that the real form of $A_{31}$ is specified by precisely the
same ingredients as the real form of $H_{11}$, namely by specifying
the hermiticity properties of $\gamma_{i,i+1} \in A_{31}$
(corresponding to $T_{\alpha_i} \in H_{11}$) and the hermiticity
properties of $\gamma_{9\,10\,11} \in A_{31}$ (which corresponds to
$T_{\alpha_0} \in H_{11}$) (compare with the arguments of
\cite{Keurentjes:2003hc}) . Furthermore, the argument on the
commutators from subsection \ref{h11} extends straightforwardly to the algebra
of $A_{31}$. We conclude that the truncation procedure must always
result in the same real form of $A_{31}$, and cannot depend on the the
choice of $A_{10}$ subalgebra of $E_{11}$. Hence the groups
$SU(32),SL(32,\R), SU^*(32)$ and $SU(16,16)$ cannot be specific to the
theories for which we have computed them, but must be the same for all
the theories connected by the Weyl group. This
assertion can of course be verified by explicit computations.  

\section{Discussion and conclusions}

The $E_{11}$ conjecture asserts that $M$-theory has a non-linearly
realized $E_{11}$ symmetry \cite{West:2001as}, described by the coset
$E_{11}/H_{11}$. The theory supposedly can be reconstructed from a
level expansion, based on a regular $A_{10}$ sub-algebra
\cite{West:2002jj, Nicolai:2003fw}. The level 0
generators are identified with the elf-bein $GL(11,R)/SO(1,10)$ of 11
dimensional general relativity. Incorporating space-times with
non-Euclidean signatures amounts to introducing signs in
$H_{11}$. There is one more sign that can be adjusted, which is the
sign in front of the 4-form term. Starting from the conventional
values for M-theory, signature $(1,10)$ and a conventional sign for the
4-form term, we have argued that the same theory should
describe Hull's $M^*$- and $M'$-theories \cite{Hull:1998ym},
essentially by choosing a different $A_{10}$ subalgebra. The sign in
front of the 4-form term in the Lagrangian found by Hull exactly
agrees with the one we find from $E_{11}$-algebra. Other signatures
than (1,10), (2,9), (5,6), (6,5), (9,2) and (10,1) are not described
by this theory: They require a different real form of the denominator
sub-algebra $H_{11}$, and therefore cannot be equivalent. We therefore
conclude that the $E_{11}$-proposal, originally set up to describe $M$-theory,
implicitly includes the $M^*$- and $M'$-theories, but no others. 

It is satisfying that a computation based on $E_{11}$ reproduces the
known theories, and no other ones. Even if not a prediction, the
existence of $M^*$-, and $M'$-theories can be regarded as a
successful, and remarkable postdiction of the conjecture. It
therefore seems that a formalism built upon a non-linearly realized
$E_{11}$-symmetry has the potential to provide a framework in which $M$-,
$M^*$- and $M'$-theory are treated as a single theory, as already
suggested by Hull. It would be interesting to see if
similar results could be extracted from other proposals for a more
fundamental description of M-theory \cite{Banks:1996vh, Horava:1997dd,
  Smolin:2000kc, Nastase:2003wb}. A superalgebra perspective on
theories with multiple time directions was offered in \cite{Bergshoeff:2000qu, Vaula:2002cn}.

We furthermore note that, although it was originally argued that these
theories are connected by time-like T-duality, in our arguments closed
time-like curves, and the doubts that one may have about them never
appear. In the $E_{11}$-context the duality simply follows from the
fact that an arbitrary choice of ``gravitational sub-algebra''
$A_{10}$ does not guarantee the space-time signature to be
$(1,10)$. If one nevertheless insists on such a space-time signature
$E_{11}$ invariance must be broken. It is easily shown that discarding
exotic space-time signatures, we must restrict to generators
corresponding to roots orthogonal to $\omega_1$, which breaks the
symmetry to $E_{10}$, and places time back in the special position that
it had lost since the advent of Relativity.  
 
As a side-remark, we note that the question of space-time signature
can be posed and answered without any resolution on the question of
how space-time, and in particular space-time translation symmetries
should be represented in the algebra \cite{Nicolai:2003fw,West:2003fc,
  Englert:2003py, Kleinschmidt:2003jf}. The possibility to add signs
for time-like directions may however be useful as a test for proposals
of higher level representations as derivatives of lower level fields
\cite{Nicolai:2003fw, Englert:2003py}. 

The techniques that were applied here to $E_{11}$, can also be applied
to theories that are dimensionally reduced over a number of directions
that includes at least one time-like one. Applied to the coset
symmetries of dimensionally reduced supergravity, they elucidate the
full duality web, and reveal duality groups that have not appeared in
the literature before. A simple link to the theory of real forms of
algebra's makes it an almost trivial exercise to classify them
\cite{Keurentjesprep}. They can also be applied to other theories
based on triple-extended algebra's \cite{Kleinschmidt:2003mf,
  Englert:2003py}, where it is easy to see that the duality pattern, and
the possible space-time signatures are very dependent on the
algebra. We intend to report on these topics in the future.   

{\bf Acknowledgements :} I would like to thank Chris Hull for correspondence.
This work was supported in part by the ``FWO-Vlaanderen'' through 
project G.0034.02, in part by the Belgian Federal Science Policy Office 
through the Interuniversity Attraction Pole P5/27 and in part by the 
European Commission RTN programme HPRN-CT-2000-00131, in which the 
author is associated to the University of Leuven.

\end{document}